\documentclass{aa}
\usepackage{amsmath}
\usepackage{graphicx}
\usepackage[varg]{txfonts}
\usepackage{natbib}
\usepackage{hyperref}
\bibpunct{(}{)}{;}{a}{}{,} % to follow the A&A style
\begin{document}
%-------------------------------------------------------------------------------
\title{Type IIP supernova 2008in: the explosion of a normal red supergiant}

\author{V. P. Utrobin\inst{1,2} \and N. N. Chugai\inst{3}}

\institute{
   Max-Planck-Institut f\"ur Astrophysik,
   Karl-Schwarzschild-Str. 1, 85748 Garching, Germany
\and
   Institute of Theoretical and Experimental Physics,
   B.~Cheremushkinskaya St. 25, 117218 Moscow, Russia
\and
   Institute of Astronomy of Russian Academy of Sciences,
   Pyatnitskaya St. 48, 119017 Moscow, Russia}

\date{Received 10 April 2013 / accepted 6 June 2013}

%===============================================================================
\abstract{% Context.
The explosion energy and the ejecta mass of a type IIP supernova make up
   the basis for the theory of explosion mechanism.
So far, these parameters have only been determined for seven events.
}{% Aims.
Type IIP supernova 2008in is another well-observed event for which a detailed
   hydrodynamic modeling can be used to derive the supernova parameters.
}{% Methods.
Hydrodynamic modeling was employed to describe the bolometric light curve
   and the expansion velocities at the photosphere level.
A time-dependent model for hydrogen ionization and excitation was applied
   to model the H$\alpha$ and H$\beta$ line profiles.   
}{% Results.
We found an ejecta mass of $13.6\pm1.9~M_{\sun}$, an explosion energy of
   $(5.05\pm3.4)\times10^{50}$ erg, a presupernova radius of
   $570\pm100~R_{\sun}$, and a radioactive $^{56}$Ni mass of
   $0.015\pm0.005~M_{\sun}$.
The estimated progenitor mass is $15.5\pm2.2~M_{\sun}$.
We uncovered a problem of the H$\alpha$ and H$\beta$ description at the early
   phase, which cannot be resolved within a spherically symmetric model.
}{% Conclusions.
The presupernova of SN~2008in was a normal red supergiant with the minimum mass
   of the progenitor among eight type IIP supernovae explored by means of
   the hydrodynamic modeling.
The problem of the absence of type IIP supernovae with the progenitor masses
   $<15~M_{\sun}$ in this sample remains open.
}
\keywords{stars: supernovae: individual: SN 2008in -- stars: supernovae: 
   general}
%-------------------------------------------------------------------------------
%
\titlerunning{Hydrodynamic model of SN 2008in}
\authorrunning{V. P. Utrobin and N. N. Chugai}
\maketitle

%===============================================================================
\section{Introduction}
\label{sec:intro}
%-------------------------------------------------------------------------------
Type IIP supernovae (SNe~IIP) comprise roughly half of core-collapse SNe
   related to massive stars \citep{SLFC_11}. 
The current theory predicts that SNe~IIP originate from $9-25~M_{\sun}$ stars
   \citep{HFWLH_03}, although the issue is still unsettled regarding the
   precise values of the lower and upper boundaries.
It is well known that the luminosities of SNe~IIP are distributed in a broad
   range, which is reflected in the categories of luminous, normal,
   subluminous, and faint events.
What does determine the difference remains unclear, although there is a hint
   that the SN luminosity may be controlled by a progenitor mass, with
   the higher luminosity events coming out from the higher stellar masses 
   \citep[see][and references therein]{UC_11}.

An understanding of SNe~IIP phenomenon requires well-determined SN parameters.
However, only the radioactive $^{56}$Ni mass can be directly estimated from
   the observed luminosity at the radioactive tail.
A detailed hydrodynamic modeling is needed to reliably recover the parameter
   values of the ejecta mass, the explosion energy, and the pre-SN radius.
In turn, the hydrodynamic modeling becomes an effective diagnostic exclusively
   in the case of the well-observed SNe~IIP, which suggests a complete light
   curve from the rising part to the radioactive tail and a detailed spectral
   coverage.
Unfortunately, only a handful of SN~IIP events met these requirements 
   which explained a scarcity of SNe~IIP with the detailed hydrodynamic modeling
   \citep[see][]{UC_11}.

Given this fact, we focus on SN~2008in, the recent SN~IIP exploded in
   nearby galaxy M61 (NGC 4303) and well observed photometrically from the
   rising part to the radioactive tail \citep{RKB_11}.
This is the second case, after SN~2005cs, of well-studied observationally
   subluminous SN~IIP. 
Using scaling relations between the observables and the SN parameters,
   \citet{RKB_11} find the ejecta mass of $\sim 16.7~M_{\sun}$, the explosion
   energy of $\sim 5.4\times10^{50}$ erg, and the pre-SN radius of
   $\sim 126~R_{\sun}$ and conclude that SN~2008in is the result of
   a low-energy explosion of a relatively compact star.
We would like to emphasize that the estimates based on the analytical
   relations between the observables (the plateau duration, the luminosity,
   and the velocity at the photosphere level in a certain epoch) and the SN
   parameters (the explosion energy, the ejecta mass, and the pre-SN radius)
   should be considered as approximate; a detailed hydrodynamic modeling of 
   the light curve and the velocity evolution is needed to infer the reliable
   values of SN parameters.    

Below we present the results of a hydrodynamic modeling of SN~2008in event,
   using the observational data reported by \citet{RKB_11}.
The results turn out to be of interest in two respects.
The derived SN parameters significantly improve earlier estimates of the ejecta
   mass and the pre-SN radius. 
Secondly, we uncover an unexpected problem of the inconsistency between the
  H$\alpha$ and H$\beta$ lines in early spectrum, which has 
  interesting implications for the structure of external ejecta.

We begin with the description of the model and the observational data of
   SN~2008in (Sect.~\ref{sec:moddat}).
We then turn to the results of the hydrodynamic modeling
   (Sect.~\ref{sec:res-snpar}) and consider the modeling of hydrogen lines
   in the early spectrum (Sect.~\ref{sec:res-hlpro}).
Section~\ref{sec:discon} summarizes the results obtained and discusses them
   in the context of other SNe~IIP. 
 
Below we adopt, following \citet{RKB_11}, the distance to SN~2008in
   $D=13.19$ Mpc and the reddening $E(B-V)=0.098$ mag.

%===============================================================================
\section{Model and observational data}
\label{sec:moddat}
%-------------------------------------------------------------------------------
%
\begin{figure}[t]
   \includegraphics[width=\hsize, clip, trim=17 152 67 64]{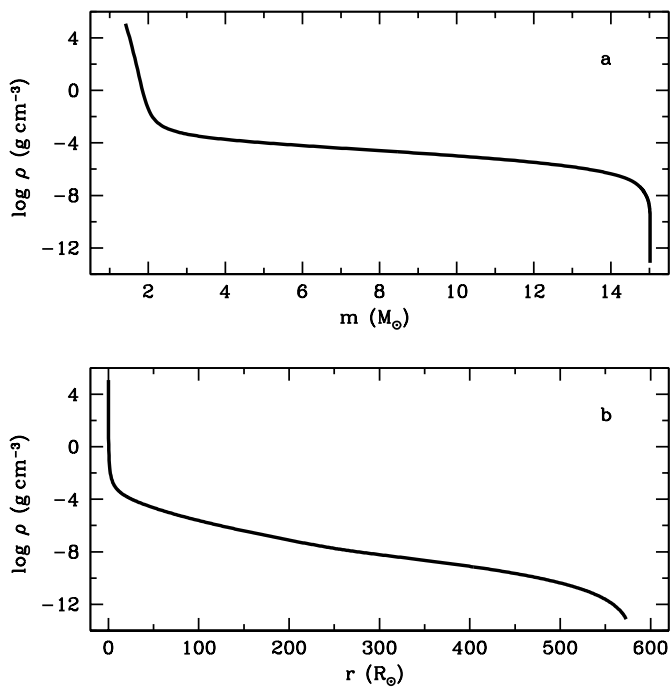}
   \caption{%
   Density distribution as a function of interior mass (Panel \textbf{a}) and
   radius (Panel \textbf{b}) for the optimal pre-SN model of SN~2008in.
   The central core of 1.4 $M_{\sun}$ is omitted.
   }
   \label{fig:denmr}
\end{figure}
The modeling of the SN explosion was performed using the spherically-symmetric
   hydrodynamic code with one-group radiation transfer \citep{Utr_04,
   Utr_07}, which has been applied previously to other SNe~IIP.
When applied to the normal type IIP SN~1999em \citep{Utr_07}, the code leads
   to the basic SN parameters similar to those recovered by \citet{BBP_05}
   who employ the hydrodynamics with multi-group radiation transfer.
This concordance suggests that the SN~IIP modeling is not hampered by
   the one-group approximation for the radiation transfer.
The basic equations and details of the input physics, including calculations of
   mean opacities, are described in \citet{Utr_04}.
The present version of the code includes additional Compton cooling and
   heating.

\begin{figure}[t]
   \includegraphics[width=\hsize, clip, trim=18 152 67 287]{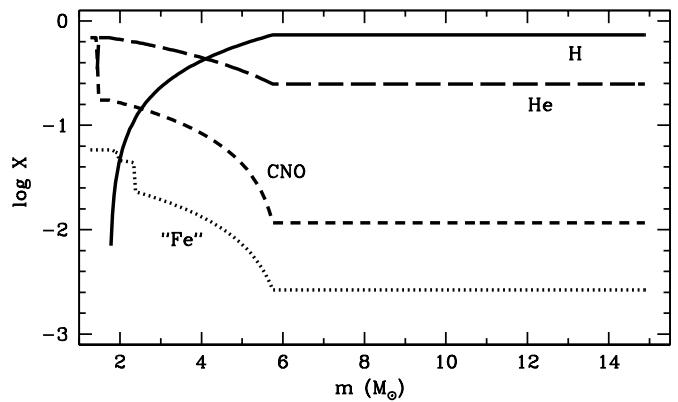}
   \caption{%
   The mass fraction of hydrogen (\emph{solid line\/}), helium
      (\emph{long dashed line\/}), CNO elements (\emph{short dashed line\/}),
      and Fe-peak elements including radioactive $^{56}$Ni
      (\emph{dotted line\/}) in the ejecta of the optimal model.
   }
   \label{fig:chcom}
\end{figure}

It is general wisdom that a normal SN~IIP originates from an explosion of a 
   massive red supergiant (RSG) star \citep{GIN_71, FA_77, EWWP_94}.
Although SN~2008in is an underluminous SN~IIP, its light curve and expansion 
   velocities leave no doubt that this event is caused by the explosion of 
   a RSG star as well. 
In order to study SN~2008in, we construct, as usual, a non-evolutionary RSG
   pre-SN model in hydrostatic equilibrium.
There are two arguments in favor of non-evolutionary pre-SN models
   \citep{UC_08}. 
Firstly, the available evolutionary pre-SN models are not able to describe the
   shape of the light curve, especially at the end of the plateau.
This problem was originally met in the case of SN~1987A \citep{Woo_88, SN_90}
   and was solved by invoking a mixing between the helium core and the hydrogen
   envelope.
Secondly, the multi-dimensional hydrodynamic simulations \citep{MFA_91,
   KPSJM_03, HJM_10} demonstrate that the SN~IIP explosion is indeed accompanied
   by mixing and smoothing of the density and composition gradients between the
   helium core and the hydrogen envelope.
Because the mixing caused by the explosion is essentially the multi-dimensional
   effect, we mimic it by a ``hand-made'' non-evolutionary RSG configuration
   adjusted to fit the observed light curve and the character of the ejecta
   expansion.
The adopted pre-SN density distribution and the chemical composition are shown
   in Figs.~\ref{fig:denmr} and \ref{fig:chcom}, respectively. 
We note that the mixing in our pre-SN model presumably reflects a combined
   effect of mixing during the stellar evolution \citep[cf.][]{HMM_04} and 
   mixing stimulated by the SN explosion.

The SN explosion is initiated by a supersonic piston applied to the bottom of
   the stellar envelope at the boundary with the $1.4~M_{\sun}$ central core,
   which is removed from the computational mass domain and assumed to collapse
   to become a neutron star.
The explosion energy we report below is the difference between the piston energy
   input and the modulus of the total energy of the envelope outside
   the collapsing core. 

In addition to modeling the SN light curves, we analyze the line profiles with
   the atmosphere model which is based on the time-dependent ionization and
   excitation kinetics of hydrogen and other elements, the time-dependent
   kinetics of molecular hydrogen, and the time-dependent energy balance
   for the gas temperature \citep{UC_05}.
The density distribution, chemical composition, radius of the photosphere,
   and effective temperature are provided by the hydrodynamic model.
The obtained time-dependent structure of the atmosphere is then used to
   model synthetic spectra at selected epochs.
The Sobolev local escape approximation is assumed for the line radiation
   transfer dominated by the line absorption.
The line emissivity and the Sobolev optical depth are determined by level
   populations provided with the time-dependent approach.
The Thomson scattering on free electrons, Rayleigh scattering on neutral
   hydrogen, and the relativistic effects are also taken into account.
Photons striking the photosphere are assumed to be diffusively reflected back
   into the atmosphere with the albedo calculated according to \citet{CU_00}.
The spectra are simulated by means of the Monte Carlo technique.

The hydrodynamic modeling with the one-group radiation transfer is aimed at
   reproducing a bolometric light curve.
For SN~2008in this light curve is recovered using $UBVRIJH$ photometry
   \citep{RKB_11} corrected for the reddening and the zero-points reported by 
   \citet{BCP_98}.
As usual, we use a black-body spectral fit, which is not a perfect procedure.
To estimate errors of the observed bolometric light curve we employ
   a black-body spectrum modified with a variable ultraviolet reduction factor
   derived from the SN~1987A spectral data \citep{PHHN_88, PKS_95}.
We find that the black-body fit for SN~2008in results in a slightly higher
   bolometric luminosity than the modified black-body fit.
The difference is less than 5\% for the first 20 days and increases up to 10\%
   at the end of the plateau.
We do not consider this difference to represent an actual error, however, because
   the SN~1987A spectra show an infrared excess in the $K$ and $L$ bands over
   the black-body approximation derived from the $BVRIH$ data \citep{CMM_87},
   which means that the real errors related to the black-body fit are likely
   to be smaller than our estimates. 

The velocity at the photosphere level is another crucial observable 
   constraining the hydrodynamic model.
We derive the photospheric velocity via fitting the spectral line profiles
   by a simple model described in Sect.~\ref{sec:res-hlpro}.
Using the H$\alpha$, H$\beta$, and He\,I 5876~\AA\ lines on day 11,
   the H$\beta$ line on day 18, and the Na\,I doublet profile on day 59, we
   find the photospheric velocity of 6020, 5150, and 2100~km\,s$^{-1}$ 
   with the uncertainty of $\pm100$ km\,s$^{-1}$ for these three epochs. 

As we will see below, our hydrodynamic model is able to fit not only the
   bolometric light curve but the $R$-band light curve as well.
This is a remarkable result, since the SN detection in the ROTSE-IIIb
   \footnote {The Robotic Optical Transient Search Experiment (ROTSE-III)
   is a set of four, 45-cm, fully robotic optical telescopes.}
   images gives the first point of the $R$-band light curve \citep{RKB_11}.
Synchronizing the calculated $R$ light curve with the first point observed 
   (Fig.~\ref{fig:blcvph}d) suggests the explosion date to be JD=2454822.0,
   nearly four days earlier than that accepted by \citeauthor{RKB_11} 
Henceforth we adopt our estimate of the explosion date.

%===============================================================================
\section{Results}
\label{sec:res}
%-------------------------------------------------------------------------------
%
%===============================================================================
\subsection{Supernova parameters}
\label{sec:res-snpar}
%-------------------------------------------------------------------------------
A search for the optimal model is facilitated by the study of parameter
   variations for the hydrodynamic model of the normal type IIP SN~1999em
   described earlier \citep{Utr_07}. 
This knowledge combined with a sample of the hydrodynamic models for the
   well-observed SNe~IIP \citep{UC_11} provide us with efficient recipes
   to search for the optimal SN parameter set. 
We note that the radioactive $^{56}$Ni mass, which is an essential parameter,
   can be recovered in a model-independent way from the bolometric luminosity
   at the radioactive tail.
This procedure results in the $^{56}$Ni mass of $0.015~M_{\sun}$ in a very
   good agreement with \citet{RKB_11}.
Another approach employs a comparison of the SN~2008in luminosity in the $R$
   band to that of SN~1987A at the same epoch.
This method gives a slightly lower $^{56}$Ni mass of $0.012~M_{\sun}$, when
   adopting the value of $0.076~M_{\sun}$ for SN~1987A \citep{Utr_05}.
Our experience with SN~2005cs \citep{UC_08}, resembling SN~2008in in the
   photometric and spectral properties, suggests that the pre-SN radius of
   SN~2008in should lie between $460$ and $740~R_{\sun}$.
The preliminary simulations of the light curve and the expansion velocities
   for SN~2008in show that the appropriate model has the ejecta mass in the
   range of $12.5-14.5~M_{\sun}$ and the explosion energy in the range of
   $(4-6)\times10^{50}$ erg.
For the detailed analysis we therefore adopt the helium core mass of
   $\sim4~M_{\sun}$, typical for the non-rotating pre-SN model of
   a $15~M_{\sun}$ main-sequence progenitor \citep{HMM_04}.
It should be noted that the light curve is not sensitive to the mass of
   the helium core; instead the light curve depends on a total ejecta
   mass and a degree of mixing between the helium core and the hydrogen
   envelope \citep{Utr_07}.

\begin{figure}[t]
   \includegraphics[width=\hsize, clip, trim=10 155 67 171]{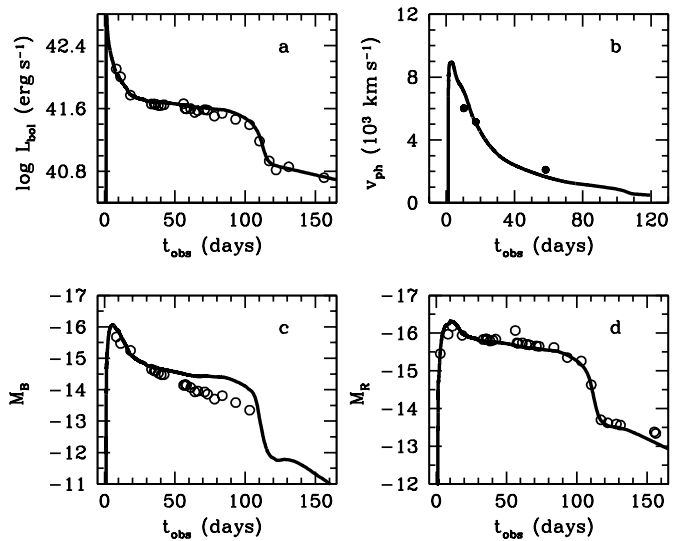}
   \caption{%
   Optimal hydrodynamic model.
   Panel \textbf{a}: the bolometric light curve of the optimal model
      (\emph{solid line\/}) overplotted on the bolometric data of SN 2008in
      (\emph{open circles\/}) evaluated from the $UBVRIJH$ magnitudes
      reported by \citet{RKB_11}.
   Panel \textbf{b}: the calculated photospheric velocity (\emph{solid
      line\/}) is compared to the photospheric velocity estimated from the 
      H$\alpha$, H$\beta$, He\,I~5876~\AA, and Na\,I doublet profiles in the
      spectra presented by \citeauthor{RKB_11} (\emph{filled circles\/}).
   Panels \textbf{c} and \textbf{d}: the calculated B and R light curves
      (\emph{solid line\/}) compared to the observations of SN 2008in
      (\emph{open circles\/}) obtained by \citeauthor{RKB_11}
   }
   \label{fig:blcvph}
\end{figure}
\begin{figure}[b]
   \includegraphics[width=\hsize, clip, trim=15 154 17 162]{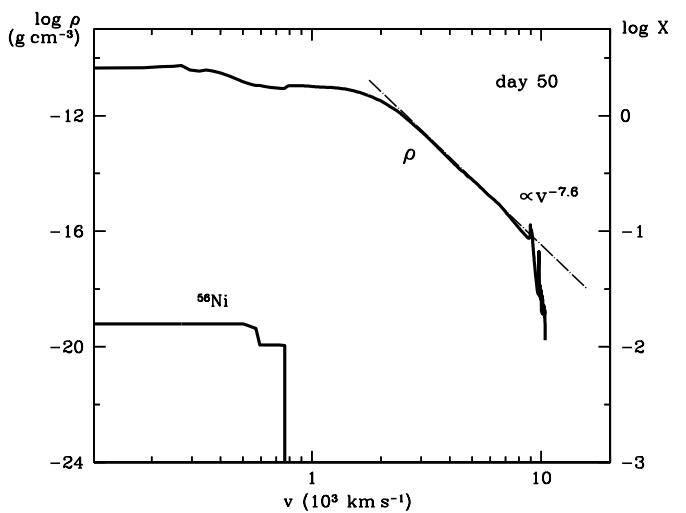}
   \caption{%
   The density and the $^{56}$Ni mass fraction as a function of velocity
      for the optimal model at $t=50$ days (\emph{solid lines\/}).
   \emph{Dash-dotted line\/} is the density distribution fit
      $\rho\propto v^{-7.6}$.
   }
   \label{fig:deni}
\end{figure}
The hydrodynamic modeling of SN~2008in for an extended parameter set led
   us to the optimal choice of the ejecta mass $M_{env}=13.6~M_{\sun}$, the
   explosion energy $E=5.05\times10^{50}$ erg, the pre-SN radius
   $R_0=570~R_{\sun}$, and the radioactive $^{56}$Ni mass
   $M_{\mathrm{Ni}}=0.015~M_{\sun}$. 
This is demonstrated by a good fit of the calculated bolometric light curve
   and the photospheric velocity to those observed (Fig.~\ref{fig:blcvph}).
The inferred ejecta mass and the explosion energy are close to the values found
   earlier by \citet{RKB_11} employing scaling relations.
However, our pre-SN radius is four times larger.
The latter suggests that the SN parameters derived from the scaling relations 
   should be cautiously accepted. 
It should be emphasized that the broad initial luminosity peak
   (Fig.~\ref{fig:blcvph}a) directly indicates the explosion of an extended
   RSG star, not a compact one.
While the overall evolution of the velocity at the photosphere level is well
   reproduced (Fig.~\ref{fig:blcvph}b), on day 11 the model velocity is 20\%
   higher than the observed value.
This mismatch may be related to the H$\alpha$ and H$\beta$ problem we discuss
   below in Sect.~\ref{sec:res-hlpro}.

The model density distribution in the freely expanding envelope on day 50
   (Fig.~\ref{fig:deni}) is similar to that of SN~2005cs \citep{UC_08}
   with the outer density power law $\rho\propto v^{-7.6}$.
The power-law index $k=-\partial \ln \rho / \partial \ln v$ depends on
   the density distribution of pre-SN outer layers, which in turn is
   constrained by the initial luminosity peak. 
The rule of thumb states that a more luminous and longer initial luminosity
   peak requires a shallower density distribution in the outer layers,
   i.e., a lower $k$ value. 
In the case of SN~2008in, the $k$ value is determined with an accuracy of
   about $\pm0.3$.   
It is worth noting that the modeling of three SNe~IIP, namely, SN~2004et
   \citep{UC_09}, SN~2005cs \citep{UC_08}, and SN~2008in, results in
   a similar density gradient with $k\approx7.6$ in the outer layers.

The principal parameters ($M_{env}$, $E$, and $R_0$) of SN~2008in are
   similar to those of SN~2005cs \citep{UC_08} (Table~\ref{tab:sumtab}),
   although the ejecta mass is somewhat lower and closer to that of the
   low-luminosity type IIP SN~2003Z \citep{UCP_07}.
The $^{56}$Ni mass in SN~2008in is higher than that of SN~2005cs by a factor
   of two.
Remarkably, the outer velocity of $^{56}$Ni material of $770$ km\,s$^{-1}$
   also exceeds that of SN~2005cs by $160$ km\,s$^{-1}$.

Combining the ejecta mass with the mass of the neutron star, we obtain
   the pre-SN mass of $15~M_{\sun}$.
This is the lower limit of a ZAMS progenitor mass because some amount
   is lost via the stellar wind.
Following our previous estimate for SN~2003Z \citep{UCP_07}, we adopt
   the amount of the lost mass to be $0.2<M_w<0.8~M_{\sun}$.
With this value of the lost mass, the progenitor mass of SN~2008in is then
   $M=15.5\pm0.3~M_{\sun}$.

The question of possible errors of the parameter value is crucial. 
We estimate the errors by calculating the hydrodynamic models with
   the parameters varied around the optimal model.
Adopting the uncertainties of the observables to be 35\% in the luminosity,
   2\% in the velocity, and 1\% in the plateau duration,
   we come to the errors in the initial radius of $\pm100~R_{\sun}$,
   the ejecta mass of $\pm1.9~M_{\sun}$, the explosion energy of
   $\pm3.4\times10^{50}$ erg, and the $^{56}$Ni mass of $\pm0.005~M_{\sun}$. 
We note that the largest error in the explosion energy is related to
   the large uncertainty of the observed luminosity.
The error of the ejecta mass combined with the uncertainty of the mass loss
   results in the error of the progenitor mass of $\pm2.2~M_{\sun}$.

%===============================================================================
\subsection{Hydrogen line problem}
\label{sec:res-hlpro}
%-------------------------------------------------------------------------------
%
\begin{figure}[t]
   \includegraphics[width=\hsize, clip, trim=27 318 67 170]{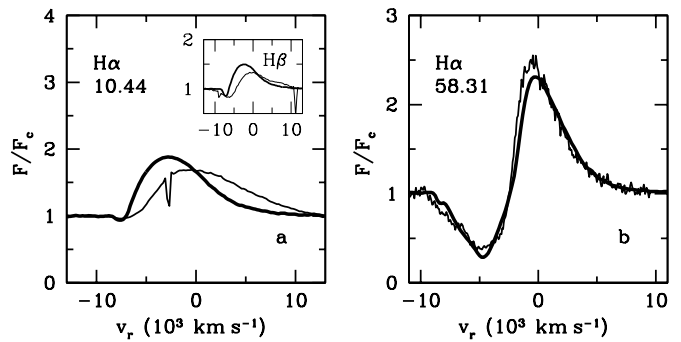}
   \caption{%
   The H$\alpha$ profile (\emph{thick solid line\/}) on days 11 and 59,
      calculated with the atmosphere model, is overplotted on the observed
      profile (\emph{thin solid line\/}) \citep{RKB_11}.
   The inset in Panel~\textbf{a} shows the H$\beta$ profiles on day 11.
   }
   \label{fig:ha}
\end{figure}
To recover the information on the external rarefied layers imprinted in the 
   H$\alpha$ wings, we solved the time-dependent atmosphere model of the hydrogen
   ionization and excitation upon the background of the optimal hydrodynamic
   model \citep[for details see][]{UC_05}.
The results obtained in this way are a little odd (Fig.~\ref{fig:ha}).
While the late H$\alpha$ profile on day 59 is reproduced satisfactorily, the 
   early profile on day 11 is very different from the observed line. 
The major drawback of the model is a pronounced blueshift, which is also
   apparent in the H$\beta$ line on day 11 (see the inset in Fig.~\ref{fig:ha}a).
The attempt to vary the hydrodynamics of the outer layers does not remove the 
   problem.

The first thought is that the uncovered mismatch could be an outcome of a large
   velocity at the photosphere level in the hydrodynamic model; this could be
   responsible for the strong occultation effect.
Indeed, we found earlier that on day 11 the empirical velocity of 
   6020 km\,s$^{-1}$ is lower than the model value of 7130 km\,s$^{-1}$
   (Fig.~\ref{fig:blcvph}b).
However, we show below that this is not the principal reason.
Here we notice that a small dip in the blue absorption wing of the model
   H$\alpha$ profile on day 59 at the radial velocity of $-8200$ km\,s$^{-1}$
   (Fig.~\ref{fig:ha}b) is the trace of the boundary shell with a mass of
   $\sim 10^{-3}~M_{\sun}$, which forms during the shock breakout
   (Fig.~\ref{fig:deni}).
A similar dip is observed in some SNe~IIP \citep{CCU_07}, but is 
   absent in SN~2008in.

\begin{figure}[t]
   \includegraphics[width=\hsize, clip, trim=20 374 67 170]{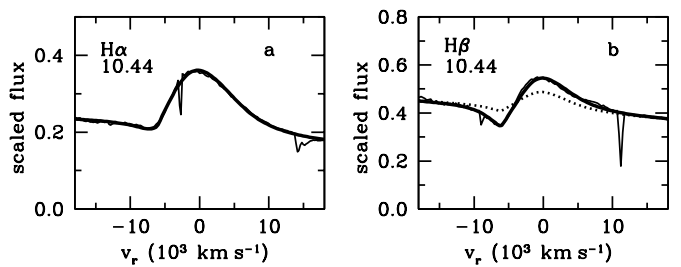}
   \caption{%
   The best-fit simulations of the H$\alpha$ and H$\beta$ lines (\emph{thick
      lines\/}) are overplotted on the corresponding profiles observed on day 11
      (\emph{thin lines\/}).
   Unlike the calculations shown in Fig.~\ref{fig:ha}, here we use a simple
      model with the radial distributions of the Sobolev optical depth and
      the line emissivity, which are similar for the H$\alpha$ and H$\beta$
      lines, but scaled arbitrarily to fit both profiles.
   \emph{The dotted line\/} represents the H$\beta$ line for the theoretical
      ratio $R_{\tau}=7.25$.
   }
   \label{fig:hafit}
\end{figure}
\begin{figure}[t]
   \includegraphics[width=\hsize, clip, trim=20 374 67 170]{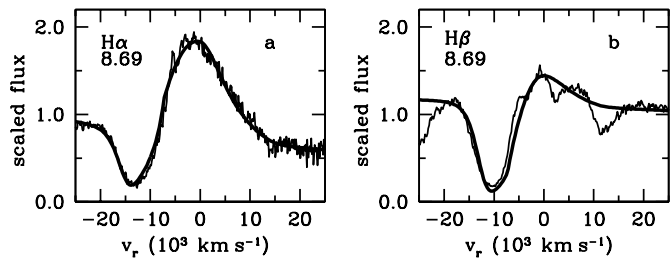}
   \caption{%
   The H$\alpha$ and H$\beta$ lines observed in SN~1987A on day 9
      (\emph{thin lines\/}) \citep{PHHN_88} and the calculated profiles
      (\emph {thick lines\/}) in the simple model with the theoretical
      ratio $R_{\tau}=7.25$.
   }
   \label{fig:hab87a}
\end{figure}

A question arises whether the difference between the model profile and 
   the observed line on day 11 is an outcome of the inadequate distribution 
   of the ejecta density, the ionization, and the excitation in the outer layers,
   or the result of some other factors.
To explore this issue, we use a parametrized model of the line
   formation, which admits a variation of the distributions of the Sobolev
   optical depth and the line emissivity in a wide range.
The ratio of the Sobolev optical depth of the H$\alpha$ and H$\beta$ lines is
   a constant determined by atomic data
   $R_{\tau}=\tau_{23}/\tau_{24}=7.25$.
This ratio can be slightly modified by the stimulated emission; this effect is
   included in the numerical computations.
A surprising result of these simulations is that both H$\alpha$ and H$\beta$
   lines cannot be described simultaneously in the framework of a spherical
   model.
The profiles can be reproduced {\em if and only if} we abandon the theoretical
   ratio $R_{\tau}=7.25$: the best fit of the H$\alpha$ and H$\beta$ profiles
   on day 11 (Fig.~\ref{fig:hafit}) is attained for the ratio $R_{\tau}=2.5$,
   three times lower.
To put it simply, the H$\alpha$ absorption component is significantly 
   weaker than expected from the strength of the H$\beta$ absorption. 
To emphasize the apparent oddity of this phenomenon, we show the
   H$\alpha$ and H$\beta$ lines in the spectra of SN~1987A on day 9
   \citep{PHHN_88} and their reasonable fit for the theoretical ratio of
   the Sobolev optical depths (Fig.~\ref{fig:hab87a}). 

We have no ready explanation for the disparity found between the H$\alpha$ and
   H$\beta$ lines.
Two suggestions are conceivable and both are related to deviations from
   the spherical symmetry.
For instance, the outer layers are admittedly clumpy.
In that case the effective optical depth of a line is determined not only by 
   the atomic cross-sections, but also by the clumpiness parameters.
For optically thick clumps, a situation is plausible when the strengths of the
   H$\alpha$ and H$\beta$ absorptions become comparable.
Thus, it imitates the case that we find in the early spectrum of SN~2008in.
Another possibility is that a large-scale emission asymmetry in the near
   hemisphere could originate from either the overall ejecta asymmetry or
   the asymmetric $^{56}$Ni ejecta, both being presumably related to
   the explosion asymmetry.
This asymmetry might produce the found disparity if the contribution of
   the emission asymmetry is significant in the H$\alpha$ line and rather weak
   in the H$\beta$ line. 
It is quite plausible given the smaller optical depth in the H$\beta$ line
   and the conversion of the H$\beta$ photons into the P$\alpha$ and H$\alpha$
   quanta. 
The H$\alpha$ emission asymmetry can fill in the absorption component, thus
   resulting in the week H$\alpha$ absorption.
We kept ourselves to these general remarks and plan to study the 
   H$\alpha$ {\em vs.} H$\beta$ disparity in detail elsewhere.

%===============================================================================
\section{Discussion and conclusions}
\label{sec:discon}
%-------------------------------------------------------------------------------
%
\begin{table}[b]
\caption[]{Hydrodynamic models of type IIP supernovae.}
\label{tab:sumtab}
\centering
\begin{tabular}{ l c c c c c c }
\hline\hline
\noalign{\smallskip}
 SN & $R_0$ & $M_{env}$ & $E$ & $M_{\mathrm{Ni}}$ 
       & $v_{\mathrm{Ni}}^{max}$ & $v_{\mathrm{H}}^{min}$ \\
       & $(R_{\sun})$ & $(M_{\sun})$ & ($10^{51}$ erg) & $(10^{-2} M_{\sun})$
       & \multicolumn{2}{c}{(km\,s$^{-1}$)}\\
\noalign{\smallskip}
\hline
\noalign{\smallskip}
 1987A &  35  & 18   & 1.5   & 7.65 &  3000 & 600 \\
1999em & 500  & 19   & 1.3   & 3.6  &  660  & 700 \\
2000cb &  35  & 22.3 & 4.4   & 8.3  &  8400 & 440 \\
 2003Z & 230  & 14   & 0.245 & 0.63 &  535  & 360 \\
2004et & 1500 & 22.9 & 2.3   & 6.8  &  1000 & 300 \\
2005cs & 600  & 15.9 & 0.41  & 0.82 &  610  & 300 \\
2008in & 570  & 13.6 & 0.505 & 1.5  &  770  & 490 \\
2009kf & 2000 & 28.1 & 21.5  & 40.0 &  7700 & 410 \\
\noalign{\smallskip}
\hline
\end{tabular}
\end{table}
The primary goal of the paper was to recover the parameters of the subluminous
   type IIP SN~2008in from the observational data of \citet{RKB_11} using
   hydrodynamic simulations.
We find the ejecta mass $M_{env}=13.6\pm1.9~M_{\sun}$, the explosion energy
   $E=(5.05\pm3.4)\times10^{50}$ erg, and the pre-SN radius
   $R_0=570\pm100~R_{\sun}$.
The $^{56}$Ni mass estimated from the radioactive tail is
   $M_{\mathrm{Ni}}=0.015\pm0.005~M_{\sun}$.
The earlier estimates of $E$ and $M_{\mathrm{Ni}}$ reported by \citet{RKB_11}
   are in reasonable agreement with our results, although our value of
   the ejecta mass is $3~M_{\sun}$ lower than that of \citeauthor{RKB_11}
More importantly, we do not confirm the earlier conclusion that the pre-SN was
   a relatively compact: our pre-SN radius of $570~R_{\sun}$ suggests that
   the pre-SN was a normal RSG star. 
Taking the mass of the neutron star and the mass loss into account, we
   estimate the progenitor mass to be $M_{\mathrm{ZAMS}}=15.5\pm2.2~M_{\sun}$. 

\begin{figure}[t]
   \includegraphics[width=\hsize, clip, trim=10 162 40 64]{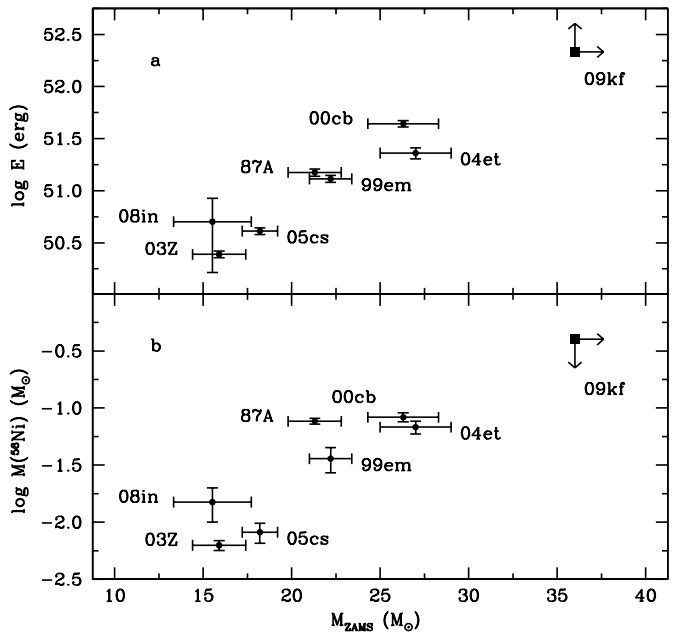}
   \caption{%
   Explosion energy (Panel \textbf{a}) and $^{56}$Ni mass (Panel \textbf{b})
      vs. hydrodynamic progenitor mass for SN~2008in and seven other
      core-collapse SNe \citep{UC_11}.
   }
   \label{fig:ennims}
\end{figure}
The parameters of SN~2008in are similar to those of another underluminous
   type IIP SN~2005cs (Table~\ref{tab:sumtab}). 
The only significant difference is that the $^{56}$Ni mass in SN~2008in is
   a factor of two higher than that of SN~2005cs.
It is a somewhat puzzling result given the higher or at least comparable ejecta
   mass of SN~2005cs.
This indicates that the explosion conditions, imprinted in the amount of
   synthesized $^{56}$Ni, are not a unique function of the progenitor mass
   of SNe~IIP. 

It is instructive to place SN~2008in on the diagrams of the explosion energy
   {\em vs.} the progenitor mass and the $^{56}$Ni mass {\em vs.} the
   progenitor mass (Fig.~\ref{fig:ennims}) together with all the rest of
   SNe~IIP studied hydrodynamically \citep{UC_11}.
We note that the error in the explosion energy for SN~2008in, which is larger
   than for other SNe~IIP, is related to the larger uncertainty in the distance
   combined with the reddening error.
The supernova 2008in falls into a broad strip occupied by SNe~IIP, thus
   confirming an assumption that the explosion energy and the $^{56}$Ni mass
   correlate with the progenitor mass $M_{\rm ZAMS}$.
The available sample of the hydrodynamically studied SNe~IIP is relatively
   scarce, and a larger number of such events is needed to confirm and
   to highlight these correlations, which are of great importance for
   constraining the explosion mechanism. 

We face an unexpected problem: the time-dependent model of the hydrogen
   ionization and the excitation, computed on the hydrodynamics background,
   fails to reproduce the H$\alpha$ and H$\beta$ lines in the SN~2008in
   spectrum on day 11. 
At first glance, this indicates that the radial structure of the outermost
   layers in SN~2008in differs substantially from the model hydrodynamic flow.
However, it turns out that the problem has deeper roots.
Detailed study shows that there is no way to reproduce
   the H$\alpha$ and H$\beta$ lines simultaneously in the framework of
   a spherically symmetric model.
We consider this as evidence that at least the outer ejecta
   ($v\ge7000$ km\,s$^{-1}$) are not spherical: a clumpiness and/or global
   asymmetry essentially affects the hydrogen line formation in the
   high-velocity layers.

The global asphericity of the H$\alpha$-emitting ejecta could be realized
   as a non-spherical pattern of the hydrogen ionization and the excitation
   produced by the asymmetric $^{56}$Ni ejecta.
The strong asymmetry of $^{56}$Ni ejecta was observed in the type IIP SN~2004dj
   \citep{CFS_05} and to a lesser extent in SN~1987A \citep{HEL_90}.
Although little can be said in detail on how this asymmetry is produced, it
   could be related to the explosion asymmetry.

The clumpiness of the ejecta is the well-known phenomenon among core-collapse
   SNe. 
Particularly, the oxygen ejecta show clearly a clumpy structure in 
   [O\,I] 6300 and 6363~\AA\ lines of the type IIP SN~1987A \citep{SDCS_91} and
   the type Ib/c SN~1985F \citep{Fil_91}.
Less apparent is the situation with the clumpiness of the outer ejecta.
At the late ($t\geq1$ yr) epoch, the spectra of the type IIb SN~1993J show
   a clumpy structure of the H$\alpha$-emitting shell \citep{FMB_94}, although
   it is not certain whether this clumpiness was produced during the SN
   outburst or the circumstellar interaction.
When studying the Cas~A
   \footnote {Cas A is the SN remnant presumably produced by the explosion
   of the type IIb SN \citep{KBU_08}.} morphology,
   \citet{Fes_01} detects the high-velocity ($\approx10^4$ km\,s$^{-1}$)
   nitrogen knots containing hydrogen.
This finding suggests that the external ejecta of the Cas~A parent SN are
   clumpy.
We therefore conclude that the global asymmetry caused by the $^{56}$Ni
   ejecta and/or the clumpiness of the outer layers could be considered
   a possible explanation of the H$\alpha$ and H$\beta$ disparity.
A detailed study is needed to resolve the issue.

To our knowledge, the H$\alpha$ and H$\beta$ disparity revealed for SN~2008in
   has not been ever mentioned for any SN~IIP and at the moment we cannot say
   whether this problem is characteristic of other SNe~IIP as well.
It is worth noting that the computed H$\alpha$ emission shows a strong blueshift
   compared to that observed in the early spectra of SN~2005cs
   \citep{DBB_08}.
We are not sure, however, whether this disparity for SN~2005cs is fatal,
   or if it could be eliminated by the appropriate tuning of a spherical model.
On the other hand, we know that the H$\alpha$ and H$\beta$ disparity
   is absent in SN~1987A.
This gives us a clue that the dissimilarity of the behavior of the
   H$\alpha$ and H$\beta$ lines in SN~2008in and SN~1987A is presumably
   related to a different structure of their pre-SNe.
Indeed, the shock breakout is likely to be accompanied by the fragmentation of
   the low-mass boundary shell.
One expects a more massive fragmented shell in the case of an exploding RSG
   star ($\sim10^{-3}~M_{\sun}$) (Sect.~\ref{sec:res-hlpro}) than for a blue
   supergiant pre-SN ($\sim10^{-6}~M_{\sun}$) \citep{IN_89}.
If the H$\alpha$ and H$\beta$ disparity is related to the clumpy structure
   of the outer ejecta, the difference in the mass of the fragmented shell
   might thus be responsible for the difference of the H$\alpha$ and
   H$\beta$ behavior in these SNe~IIP.

The ejecta mass of SN~2008in is the smallest among those derived by the
   hydrodynamic modeling.
Even with the conservative estimate of the progenitor mass of
   $15.5\pm2.2~M_{\sun}$, we face a challenging problem: why does not the sample
   of the well-observed SNe~IIP include the events with masses
   $\leq15~M_{\sun}$?
This problem has already been discussed by \citet{UC_08} and is summarized
   as follows:  either the observed sample is biased towards the luminous,
   high-mass SNe~IIP or the hydrodynamic masses are overestimated for an
   unknown reason.
At this stage both explanations seem plausible which leaves open the question
   of the progenitor mass range recovered for SNe~IIP by hydrodynamic
   modeling.

%===============================================================================
\begin{acknowledgements}
%-------------------------------------------------------------------------------
We thank Rupak Roy for kindly sending us spectra of SN~2008in.
V.P.U. is grateful to Wolfgang Hillebrandt for the possibility of working
   at the MPA.
We also thank the anonymous referee for critical comments which helped
   improve the manuscript.
%-------------------------------------------------------------------------------
\end{acknowledgements}
%-------------------------------------------------------------------------------

%===============================================================================

%-------------------------------------------------------------------------------

%-------------------------------------------------------------------------------

\begin{thebibliography}{36}
\expandafter\ifx\csname natexlab\endcsname\relax\def\natexlab#1{#1}\fi

\bibitem[{{Baklanov} {et~al.}(2005){Baklanov}, {Blinnikov}, \&
  {Pavlyuk}}]{BBP_05}
{Baklanov}, P.~V., {Blinnikov}, S.~I., \& {Pavlyuk}, N.~N. 2005, Astronomy
  Letters, 31, 429

\bibitem[{{Bessell} {et~al.}(1998){Bessell}, {Castelli}, \& {Plez}}]{BCP_98}
{Bessell}, M.~S., {Castelli}, F., \& {Plez}, B. 1998, \aap, 333, 231

\bibitem[{{Catchpole} {et~al.}(1987){Catchpole}, {Menzies}, {Monk}, {Wargau},
  {Pollaco}, {Carter}, {Whitelock}, {Marang}, {Laney}, {Balona}, {Feast},
  {Lloyd Evans}, {Sekiguchi}, {Laing}, {Kilkenny}, {Spencer Jones}, {Roberts},
  {Cousins}, {van Vuuren}, \& {Winkler}}]{CMM_87}
{Catchpole}, R.~M., {Menzies}, J.~W., {Monk}, A.~S., {et~al.} 1987, \mnras,
  229, 15P

\bibitem[{{Chugai} {et~al.}(2007){Chugai}, {Chevalier}, \& {Utrobin}}]{CCU_07}
{Chugai}, N.~N., {Chevalier}, R.~A., \& {Utrobin}, V.~P. 2007, \apj, 662, 1136

\bibitem[{{Chugai} {et~al.}(2005){Chugai}, {Fabrika}, {Sholukhova},
  {Goranskij}, {Abolmasov}, \& {Vlasyuk}}]{CFS_05}
{Chugai}, N.~N., {Fabrika}, S.~N., {Sholukhova}, O.~N., {et~al.} 2005,
  Astronomy Letters, 31, 792

\bibitem[{{Chugai} \& {Utrobin}(2000)}]{CU_00}
{Chugai}, N.~N. \& {Utrobin}, V.~P. 2000, \aap, 354, 557

\bibitem[{{Dessart} {et~al.}(2008){Dessart}, {Blondin}, {Brown}, {Hicken},
  {Hillier}, {Holland}, {Immler}, {Kirshner}, {Milne}, {Modjaz}, \&
  {Roming}}]{DBB_08}
{Dessart}, L., {Blondin}, S., {Brown}, P.~J., {et~al.} 2008, \apj, 675, 644

\bibitem[{{Eastman} {et~al.}(1994){Eastman}, {Woosley}, {Weaver}, \&
  {Pinto}}]{EWWP_94}
{Eastman}, R.~G., {Woosley}, S.~E., {Weaver}, T.~A., \& {Pinto}, P.~A. 1994,
  \apj, 430, 300

\bibitem[{{Falk} \& {Arnett}(1977)}]{FA_77}
{Falk}, S.~W. \& {Arnett}, W.~D. 1977, \aaps, 33, 515

\bibitem[{{Fesen}(2001)}]{Fes_01}
{Fesen}, R.~A. 2001, \apjs, 133, 161

\bibitem[{{Filippenko}(1991)}]{Fil_91}
{Filippenko}, A.~V. 1991, in Supernovae, ed. S.~E. {Woosley}, 467

\bibitem[{{Filippenko} {et~al.}(1994){Filippenko}, {Matheson}, \&
  {Barth}}]{FMB_94}
{Filippenko}, A.~V., {Matheson}, T., \& {Barth}, A.~J. 1994, \aj, 108, 2220

\bibitem[{{Grassberg} {et~al.}(1971){Grassberg}, {Imshennik}, \&
  {Nadyozhin}}]{GIN_71}
{Grassberg}, E.~K., {Imshennik}, V.~S., \& {Nadyozhin}, D.~K. 1971, \apss, 10,
  28

\bibitem[{{Haas} {et~al.}(1990){Haas}, {Erickson}, {Lord}, {Hollenbach},
  {Colgan}, \& {Burton}}]{HEL_90}
{Haas}, M.~R., {Erickson}, E.~F., {Lord}, S.~D., {et~al.} 1990, \apj, 360, 257

\bibitem[{{Hammer} {et~al.}(2010){Hammer}, {Janka}, \& {M{\"u}ller}}]{HJM_10}
{Hammer}, N.~J., {Janka}, H.-T., \& {M{\"u}ller}, E. 2010, \apj, 714, 1371

\bibitem[{{Heger} {et~al.}(2003){Heger}, {Fryer}, {Woosley}, {Langer}, \&
  {Hartmann}}]{HFWLH_03}
{Heger}, A., {Fryer}, C.~L., {Woosley}, S.~E., {Langer}, N., \& {Hartmann},
  D.~H. 2003, \apj, 591, 288

\bibitem[{{Hirschi} {et~al.}(2004){Hirschi}, {Meynet}, \& {Maeder}}]{HMM_04}
{Hirschi}, R., {Meynet}, G., \& {Maeder}, A. 2004, \aap, 425, 649

\bibitem[{{Imshennik} \& {Nadezhin}(1989)}]{IN_89}
{Imshennik}, V.~S. \& {Nadezhin}, D.~K. 1989, Astrophysics and Space Physics
  Reviews, 8, 1

\bibitem[{{Kifonidis} {et~al.}(2003){Kifonidis}, {Plewa}, {Janka}, \&
  {M{\"u}ller}}]{KPSJM_03}
{Kifonidis}, K., {Plewa}, T., {Janka}, H.-T., \& {M{\"u}ller}, E. 2003, \aap,
  408, 621

\bibitem[{{Krause} {et~al.}(2008){Krause}, {Birkmann}, {Usuda}, {Hattori},
  {Goto}, {Rieke}, \& {Misselt}}]{KBU_08}
{Krause}, O., {Birkmann}, S.~M., {Usuda}, T., {et~al.} 2008, Science, 320, 1195

\bibitem[{{M{\"u}ller} {et~al.}(1991){M{\"u}ller}, {Fryxell}, \&
  {Arnett}}]{MFA_91}
{M{\"u}ller}, E., {Fryxell}, B., \& {Arnett}, D. 1991, \aap, 251, 505

\bibitem[{{Phillips} {et~al.}(1988){Phillips}, {Heathcote}, {Hamuy}, \&
  {Navarrete}}]{PHHN_88}
{Phillips}, M.~M., {Heathcote}, S.~R., {Hamuy}, M., \& {Navarrete}, M. 1988,
  \aj, 95, 1087

\bibitem[{{Pun} {et~al.}(1995){Pun}, {Kirshner}, {Sonneborn}, {Challis},
  {Nassiopoulos}, {Arquilla}, {Crenshaw}, {Shrader}, {Teays}, {Cassatella},
  {Gilmozzi}, {Talavera}, {Wamsteker}, {Fransson}, \& {Panagia}}]{PKS_95}
{Pun}, C.~S.~J., {Kirshner}, R.~P., {Sonneborn}, G., {et~al.} 1995, \apjs, 99,
  223

\bibitem[{{Roy} {et~al.}(2011){Roy}, {Kumar}, {Benetti}, {Pastorello}, {Yuan},
  {Brown}, {Immler}, {Fatkhullin}, {Moskvitin}, {Maund}, {Akerlof}, {Wheeler},
  {Sokolov}, {Quimby}, {Bufano}, {Kumar}, {Misra}, {Pandey}, {Elias-Rosa},
  {Roming}, \& {Sagar}}]{RKB_11}
{Roy}, R., {Kumar}, B., {Benetti}, S., {et~al.} 2011, \apj, 736, 76

\bibitem[{{Shigeyama} \& {Nomoto}(1990)}]{SN_90}
{Shigeyama}, T. \& {Nomoto}, K. 1990, \apj, 360, 242

\bibitem[{{Smith} {et~al.}(2011){Smith}, {Li}, {Filippenko}, \&
  {Chornock}}]{SLFC_11}
{Smith}, N., {Li}, W., {Filippenko}, A.~V., \& {Chornock}, R. 2011, \mnras,
  412, 1522

\bibitem[{{Stathakis} {et~al.}(1991){Stathakis}, {Dopita}, {Cannon}, \&
  {Sadler}}]{SDCS_91}
{Stathakis}, R.~A., {Dopita}, M.~A., {Cannon}, R.~D., \& {Sadler}, E.~M. 1991,
  in Supernovae, ed. S.~E. {Woosley}, 95

\bibitem[{{Utrobin}(2004)}]{Utr_04}
{Utrobin}, V.~P. 2004, Astronomy Letters, 30, 293

\bibitem[{{Utrobin}(2005)}]{Utr_05}
{Utrobin}, V.~P. 2005, Astronomy Letters, 31, 806

\bibitem[{{Utrobin}(2007)}]{Utr_07}
{Utrobin}, V.~P. 2007, \aap, 461, 233

\bibitem[{{Utrobin} \& {Chugai}(2005)}]{UC_05}
{Utrobin}, V.~P. \& {Chugai}, N.~N. 2005, \aap, 441, 271

\bibitem[{{Utrobin} \& {Chugai}(2008)}]{UC_08}
{Utrobin}, V.~P. \& {Chugai}, N.~N. 2008, \aap, 491, 507

\bibitem[{{Utrobin} \& {Chugai}(2009)}]{UC_09}
{Utrobin}, V.~P. \& {Chugai}, N.~N. 2009, \aap, 506, 829

\bibitem[{{Utrobin} \& {Chugai}(2011)}]{UC_11}
{Utrobin}, V.~P. \& {Chugai}, N.~N. 2011, \aap, 532, A100

\bibitem[{{Utrobin} {et~al.}(2007){Utrobin}, {Chugai}, \&
  {Pastorello}}]{UCP_07}
{Utrobin}, V.~P., {Chugai}, N.~N., \& {Pastorello}, A. 2007, \aap, 475, 973

\bibitem[{{Woosley}(1988)}]{Woo_88}
{Woosley}, S.~E. 1988, \apj, 330, 218

\end{thebibliography}
\end{document}